\documentclass[%
 reprint,
 amsmath,amssymb,
 aps, physrev,
]{revtex4-2}

\usepackage{graphicx}% Include figure files
\usepackage{dcolumn}% Align table columns on decimal point
\usepackage{bm}% bold math
\usepackage{xcolor}
\usepackage[dvipsnames]{xcolor}

\begin{document}

\preprint{APS/123-QED}

\title{\textbf{Aerophilic Debubbling} 
}% 

\author{Bert J.C. Vandereydt}
\thanks{These authors contributed equally to this work.}
\author{Saurabh Nath}
\thanks{These authors contributed equally to this work.}
\author{Kripa K. Varanasi}
\email{Contact author: varanasi@mit.edu}
\affiliation{
 Department of Mechanical Engineering, Massachusetts Institute of Technology, Cambridge, Massachusetts 02139, USA
}%

\date{October 10, 2024}% It is always \today, today,
             %  but any date may be explicitly specified

\begin{abstract}

In this letter, we characterize quantitatively the complex phenomenon of debubbling via aerophilic membranes by examining local interactions at the scale of single bubbles. We identify three asymptotic limits of evacuation dictated by Rayleigh, Ohnesorge and Darcy dynamics, the physics of which we capture using simple scaling laws. We show that beyond a threshold permeability, bubble evacuations become constant in time – a feature we understand as an inertio-capillary limit. Our experiments reveal that the fastest bubble evacuations require an interface that is nearly a liquid, but not quite.

%In this letter, we experimentally show that highly permeable aerophilic membranes, when placed on a liquid, “rigidify” the liquid–air interface and annihilate bubbles within milliseconds. This ultrafast regime appears only above a critical permeability threshold, where the flow departs from classical Darcy-driven dynamics in micropores. We quantitatively characterize this aerophilicity-mediated debubbling process by examining local interactions on the scale of single bubbles approaching the membrane and identify three asymptotic evacuation regimes, the physics of which we capture through simple scaling laws.

\end{abstract}

%\keywords{Suggested keywords}%Use showkeys class option if keyword
                              %display desired
\maketitle

\footnotetext{These authors contributed equally to this work.}

%\tableofcontents

The discovery of air layers sustained within micro-textured surfaces dates back to the early 1910s, in Frank Brocher's report on aquatic insects as Haemonia mutica \cite{1-brocher1912reserches,2-thorpe1947studies}. These remarkable creatures possess a thin layer of air, now known as a plastron, trapped within the intricate hair-like structures rendering their bodies aerophilic \cite{3-flynn2008underwater, 4-balmert2011dry}. The nature of such aerophilicity – nearly a century later – is being explored in detail, including its interactions with bubbles and phenomena such as spreading and friction \cite{5-wang2009air, 6-de2016spreading, 7-rapoport2020capturing, naturecommnewpaper}, while also focusing on their design, regeneration, and stability \cite{8-patankar2016thermodynamics, 9-panchanathan2018plastron,10-wong2022effervescence, 11-tesler2023long}.

%\begin{figure*}
%\includegraphics[width =0.9\textwidth]{Fig1altV9.pdf}% Here is how to import EPS art
%\caption{\label{fig:wide} \textbf{Cascade of Scales:}\,\, (a) Schematic and timelapse of Mode I showing pore-by-pore (yellow dotted line, $\Delta t$ = 0.5 ms and 1 ms), seen in Videos 3 and 4. (c) Contact line evolution for four permeabilities, revealing Mode II as rate-limiting step. (d) Evacuation times vs. permeability for H$_2$ bubbles in water (squares, black), air in water (circles, dark red) and water/glycerol mixtures (circles, shades of red). The gray region shows a gas viscosity-limited regime, where both air and H$_2$ bubble data follow $\kappa^{-1}$ scaling, differing only by the viscosity ratio. This scaling continues until a critical permeability $\kappa_c$, above which evacuation times become constant at $\tau = 4.7 \pm 0.6$ ms. Triangular marker: capillary case , $d=R, \Phi = 1$ (Video 5). A complementary volume-time plot is in Section 3 of the SI.}
%\end{figure*}

Such explorations have recently proposed a promising material innovation: perforating an aerophilic solid to enhance bubble evacuation \cite{koreanmembrane} - crucial for CO$_2$ capture and separation \cite{12-massen2024engineering}, foam mitigation in bioreactors \cite{7-rapoport2020capturing,13-hegner2021ultrafast,wong2021weak}, methane capture from water bodies \cite{shakhova2010extensive, 15-aben2017cross, chen2012terminating}, and applications in microfluidics \cite{16-skelley2008active, 17-yang2022emergence}. Here, we explore the complex phenomenon of debubbling through porous aerophilic membranes by investigating the local microphysics of single-bubble interactions with an aerophilic interface, which helps illuminate the fundamental mechanisms that govern evacuation dynamics.

We illustrate the potential of \textit{aerophilic debubbling} in Fig 1a. A typical bioreactor sparger (Applikon Fermentation/Bioreactor Sparger, Tip 10 mm) generates bubble populations underwater that remain at the air-water interface for 1-100 s, even in the absence of surfactants. An aerophilic membrane at the interface erases such accumulations near-instantaneously, reducing evacuation times on the scale of single bubbles to milliseconds. 

To understand the fundamental mechanism driving this process, we devise a typical experiment comprising attaching the aerophilic membrane to the bottom of a tubular capture device and placing at the liquid-air interface of a water-filled tank (Fig 1a). Millimetric air bubbles of radius $R = 2$ mm released from a few centimeters below rise to meet the membrane. Membrane permeability is varied by four orders of magnitude: $0.8\, \mu$m$^2 < \kappa =d^2 \Phi <10\,\mu$m$^2$. High-permeability membranes are fabricated via photolithography on $200 \mu$m silicon wafers (pore diameter $10-200 \mu$m); low- permeability ($<10\,\mu$m$^2$) cases use calibrated PTFE membranes. Aerophilicity is achieved by dip-coating membranes in hydrophobized silica nanobeads ($\sim$30 nm, Glaco Mirror Coat), drying, and heating at 80 °C for 30 min. The process is repeated thrice for a uniform roughness of 150 nm  (AFM, Fig. S1, Section 1, SI), enabling a stable plastron of comparable thickness underwater. Advancing and receding angles are $\theta_a=142.9^\circ \pm 5.5^\circ$ (maximum) and receding as $\theta_r=97.1^\circ \pm 6.8^\circ$ (minimum) (Section 2, SI). High-speed imaging (20,000 FPS) is performed in a back-lit water tank with vertical walls to minimize distortion.
 
As the bubble approaches the membrane, a thin water film separates it from the plastron; the building lubrication pressure bounces the bubble off. Typically, a couple of bounces later ($\sim 10$ms), the thin water film separating the bubble from plastron ruptures, initiating evacuation. Fig 1b shows a timelapse of a 2 mm bubble evacuating through an aerophilic membrane with $\kappa = 78.5 \mu m^2$. Three distinct dynamical events are observed: I) at $t=0$, the ruptured water film starts dewetting across the aerophilic membrane, until it pins at $\tau_I = 1.5 ms$; II) the bubble starts evacuating with a pinned contact line, losing $\sim 95 \%$ of its volume until $\tau_I + \tau_{II} = 6$ ms; and III) contact line depins and the remnant drains in $\tau_{III} = 0.75\,$ms via stick–slip motion of the contact line. 

\begin{figure*}
\includegraphics[width =0.9\textwidth]{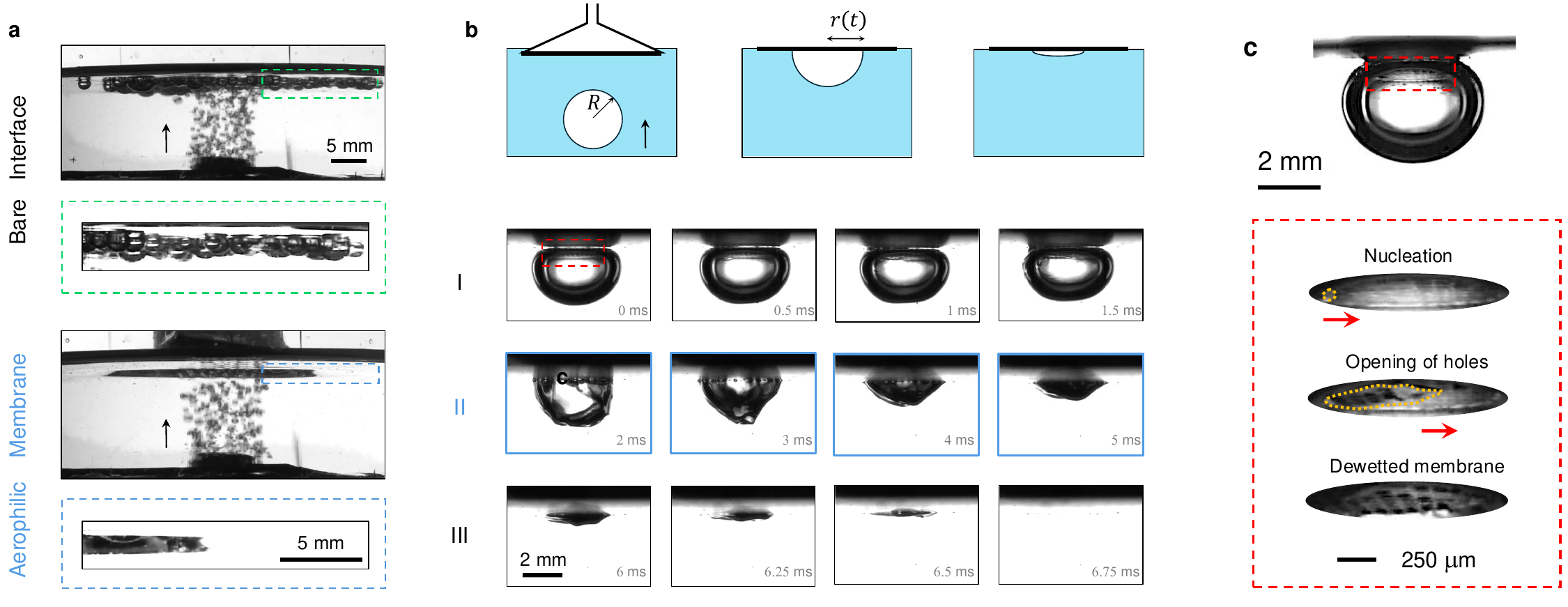}% Here is how to import EPS art
\caption{\label{fig:wide} \textbf{Cascade of Scales in Aerophilic Debubbling:}\,\, (a) \textit{Global scale dynamics:} An air sparger generates bubbles underwater at $20$\,ml/min, accumulating at the liquid-air interface for 1–100s (inset), while an aerophilic membrane-laden interface shows no bubble accumulation (inset) due to evacuations occurring within milliseconds (Video 1 in Supplemental Materials). (b) \textit{Single bubble dynamics:} Schematic and timelapse of a 2 mm radius underwater air bubble evacuating through an aerophilic membrane ($\kappa = 78.5 \mu m^2$) showing three distinct processes: I) water film dewetting, II) constant contact line evacuation (dominant evacuation, highlighted in blue) and III) depinning and evacuation. c. Pore scale dynamics: Timelapse of Mode I demonstrating pore-by-pore evacuation as the thin water film separating the bubble from the aerophilic interface dewets, marked by the yellow dotted line ($\Delta t = 0.5$ ms and $1$ ms), as seen in Videos 2 and 3 in the Supplemental Materials.}
\end{figure*}

%A typical experiment attaches the aerophilic membrane to the base of a tubular capture device at the liquid–air interface of a water-filled tank (Fig. 1a). Air bubbles of radius $R = 2$ mm released from a few centimeters below rise to meet the membrane. Once the thin film of water separating the membrane and the bubble ruptures, a three-stage evacuation process begins (Fig. 1b): 

%Fig. 2c shows the contact-line evolution for membranes of varying permeability, revealing Mode II as the slowest and therefore rate-limiting stage ($\tau \sim \tau_{II}$), a condition persisting even when the external viscosity is increased by a factor of $\sim 10^3$.

\begin{figure}[htbp]
\includegraphics[width = 0.44\textwidth]{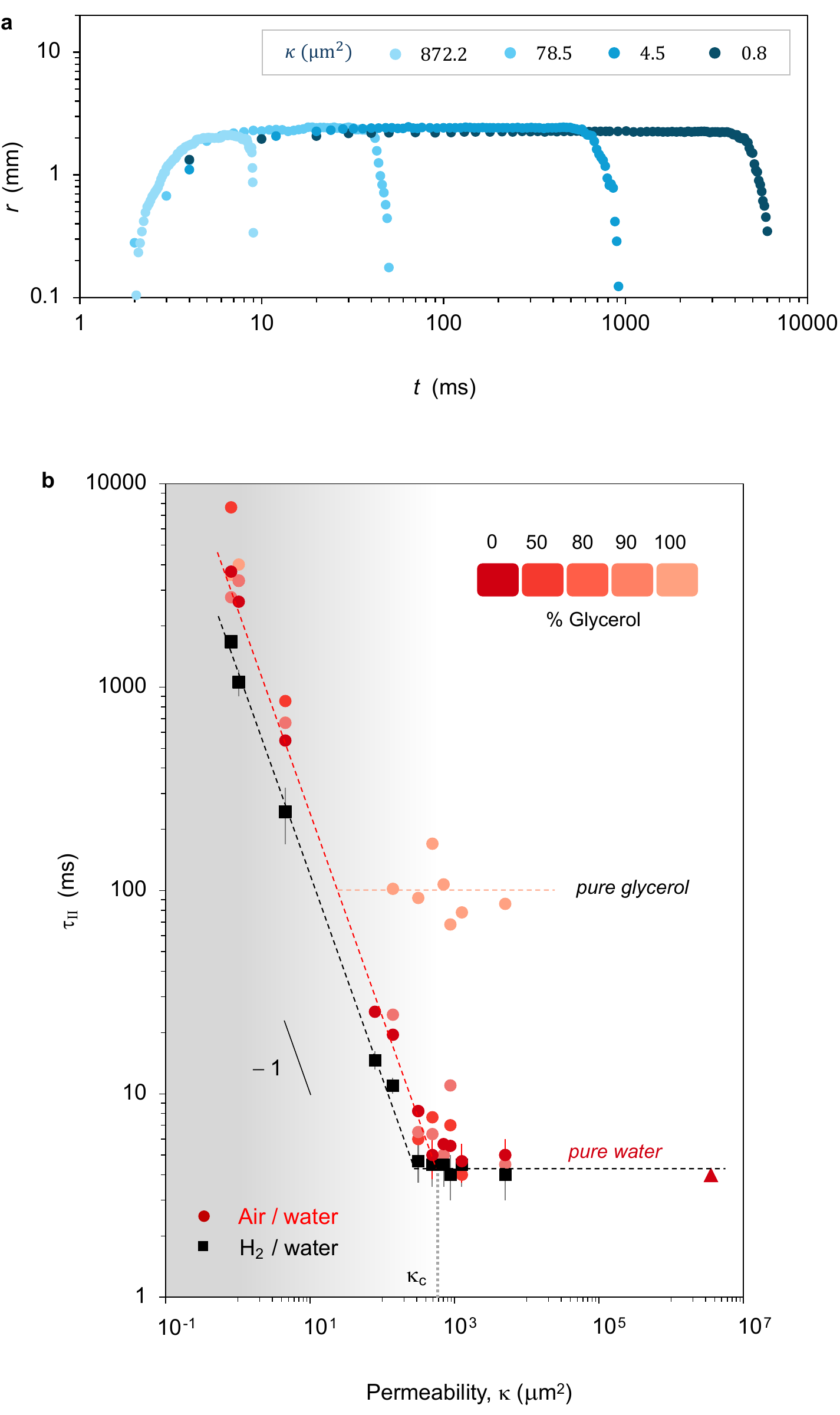}% Here is how to import EPS art
\caption{\label{fig:wide} \textbf{Evacuation times} (a) Temporal evolution of the bubble contact line for evacuation through four different permeabilities, revealing the constant contact line mode (Mode II) as the rate-limiting step (Video 2 and 4). (b) Evacuation times vs. permeability for H$_2$ bubbles in water (squares, black), air in water (circles, dark red) and water/glycerol mixtures (circles, shades of red). The gray region shows a gas viscosity-limited regime, where both air and H$_2$ bubble data follow $\kappa^{-1}$ scaling, differing only by the viscosity ratio. This scaling continues until a critical permeability $\kappa_c$, above which evacuation times become constant at $\tau = 4.7 \pm 0.6$ ms. Triangular marker: capillary case , $d=R, \Phi = 1$ (Video 5). A complementary volume-time plot is in Section 3 of the SI. }
\end{figure}

\textbf{Mode I.} 
High-speed time-lapse of Mode I in water shows a pore-by-pore opening of micropores as the water film ruptures and a dewetting front (yellow dotted line) sweeps across the membrane at $V_f \approx 2.0 \pm 0.1$ m/s (Fig.\,2b, Fig.\,S6). Dark patches mark opened pores, trailing a rim of accumulated water until $t=0.75$ ms, when the front surges ahead, leaving a cascade of dark patches. The absence of new patches indicates residual water trapped in micropores ($\sim$100 pL), by pinning defects. At 2 ms, new patches appear as water-filled pores open, zipping array-by-array at $\sim 10$-$100$ \textmu s per pore.

Mode I exhibits multi-scale dynamics: a global dewetting front and local pore-scale evacuation. The front retracts at inertial speeds ($\sim$1 m/s) consistent with Taylor–Culick dynamics, where  $V_f \sim (2\gamma/\rho e)^{1/2} \sim 1$ m/s, from a balance of surface tension $2\gamma$ and rim inertia $\rho e V_f^2$, where $\rho$ is the density of water and  $e$ the film thickness \cite{18-culick1960comments, 19-taylor1959dynamics}. Film thickness from backlit high-speed imaging (following the methodology in \cite{gauthier2016aerodynamic}) gives $e \approx 25.5 \pm  5.8 \mu $m, yielding a dewetting speed $V_f \sim (2\gamma/\rho e)^{1/2} \sim 1$ m/s. Locally, pores open in an array-by-array, zipping-like manner reminiscent of the Cassie-to-Wenzel transition \cite{20-sbragaglia2007spontaneous}, but here evacuations are inertial ($Re \sim 10^3$), with $\rho (L/\tau_{\text{pore}})^2 \sim \gamma / R$, giving $\tau_{\text{pore}} \sim 10\ \mu\text{s}$ as a local pore evacuation timescale (Section 4, SI).

When viscosified with water–glycerol mixtures, Taylor–Culick dynamics persist up to 100 mPa·s, beyond which retraction slows and pinning dominates (Fig. S3). Pinning occurs in all cases, with defect density (holes) increasing with permeability and enhancing contact-angle hysteresis (Fig. S2). Mode I thus sets the contact-line boundary for Mode II and remains rapid ($\tau_{I} \lesssim \tau_{II}$ (Fig. S3)), both behaviors being independent of permeability, viscosity (gas or liquid), or bubble size.

\textbf{Mode II (Evacuation time).} Once the retracting film reaches $r \sim R$  and pins, the Laplace pressure gradient $\gamma/R$ drives evacuation through opened pores. Increasing pore size and density (effectively permeability $\kappa = d^2 \Phi$) shortens evacuation times over three decades, with $\tau \propto \kappa^{-1}$ from seconds to milliseconds (Fig. 2d). Beyond a critical $\kappa_c \approx 500\,\mu$m$^2$, evacuation times saturate at  $\tau_c = 4.7 \pm 0.6 $ms. The same two regimes occur with H$_2$: at high $\kappa$, data for H$_2$ and air collapse on $\tau_c$; at low $\kappa$, the $\kappa^{-1}$ scaling is preserved but with the coefficient halved, matching the H$_2$-to-air viscosity ratio. These results show that below $\kappa_c$, evacuation times are set by the gas viscosity $\eta_g$. This is further supported by glycerol experiments. Despite its $\sim 10^3$ fold higher liquid viscosity $\eta$, low $\kappa$ air bubble evacuation times collapse onto the air–water data, confirming that this regime is independent of $\eta$.

\begin{figure*}
\includegraphics[width =  0.92\textwidth ]{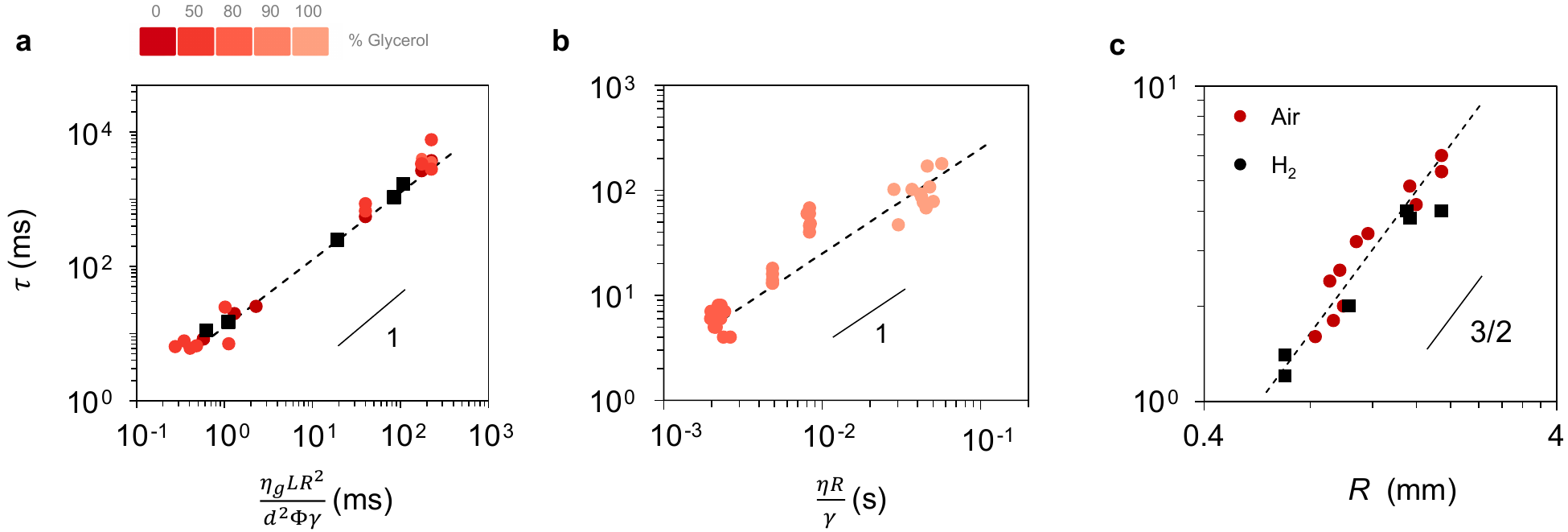}% Here is how to import EPS art
\caption{\label{fig:epsart} \textbf{Interplay of Timescales} (a) Evacuation times plotted against the Darcy timescale for $\kappa <500\,\mu$m$^2$ and water-glycerol mixtures up to 80\% glycerol by mass. Dashed line follows Eq. 1 with a pre-factor 2.4. (b) Beyond 80\% glycerol, evacuation times become proportional to the visco-capillary timescale (of the external media) with a coefficient of 2.5. (c)  Evacuation times of air and H$_2$ bubbles in water plotted against their radii reveals a 3/2 slope. The dashed line follows $\tau =2.3 (\rho R^3/\gamma)^{1/2}$.}
\end{figure*}

Gas viscous dissipation can occur in the spherical bubble cap or in flow through the pores. For the cap, balancing Stokesian dissipation $\eta_g U^2 R$ with Laplace pressure work (per unit time) $\gamma RU$ where $U \sim R/\tau $ gives $\eta_g R/\gamma$, a time independent of permeability and of the order of $0.1 \mu$s – two facts inconsistent with our low $\kappa$ data, which show strong $\kappa$ dependence and $\tau$ larger by four orders. Dissipation must therefore occur in the pores. Let 
$V$ be the gas velocity through the pores; balancing flow rates gives $U \sim V \Phi$
, with pore count $n \sim \Phi R^2/ d^2$. The viscous dissipation rate is then $\eta_g V^2 Ln$. In the adiabatic limit, balancing this with Laplace work yields
\begin{eqnarray}
\tau_{\eta_g} \sim {\eta_g LR^2}/{d^2 \Phi \gamma},
\end{eqnarray}
the classical Darcy time, proportional to $\kappa^{-1}$ and to $\eta_g$, matching Fig. 2d. Varying $\eta_g$ with air and H$_2$ and $\eta$ with water or glycerol mixtures over four decades shows that for $\kappa < \kappa_c$ all bubble evacuations follow the Darcy time with a coefficient 2.4 (Fig 3a). 

Fig. 2d shows that for mixtures above $80\%$ glycerol by mass, viscous effects in the external medium begin to dominate. In pure glycerol (lightest red in Fig. 2d), the plateau time reaches 100 ms, about 25 times slower than in water. A transition from gas viscous to liquid viscous behavior emerges with increasing permeability, the onset occurring at a much lower critical value $\kappa_c \sim 2 \mu$m$^2$. In this second viscous limit, the Laplace work $\gamma RU$ is entirely dissipated in the liquid as $\eta U^2 R$ yielding
%;  \textcolor{RoyalBlue}{substituting $U \sim R/\tau$}, we get
\begin{eqnarray}
\tau_{\eta} \sim \eta R / \gamma,
\end{eqnarray}
an Ohnesorge timescale independent of membrane porosity, consistent with the plateau in Fig. 2d. Systematic variation of liquid viscosity from $1000$mPa$\cdot $s$>\eta > 100 $mPa$\cdot$s, and bubble radii from $2.9$\,mm $> R > 0.7$\,mm, for  $\kappa > \kappa_c $ shows evacuation times scaling directly with $\tau_\eta$, with a prefactor of 2.5 (Fig. 3b).

The final regime in Fig. 2d represents the fastest possible evacuations, attained when $\eta < 100 $mPa$\cdot$s and $\kappa > \kappa_c $. Here $\tau$ is independent of gas viscosity as well as permeability, as confirmed by  increasing $\kappa$ from $10^2$ to $10^7 \mu$m$^2$, the latter corresponding to a no-solid limit: $\Phi \approx 1$, $d = R$ (a capillary tube, Video 5). This too has no effect on $\tau$ whatsoever (red triangle, Fig 2) which implies that with minimal dissipative forces, the only possibility of resistance comes from inertia. Balancing inertia $\rho (R/ \tau )^2$ with curvature pressure gives
\begin{eqnarray}
\tau_i \sim (\rho R^3/ \gamma)^{1/2},
\end{eqnarray}
the Rayleigh timescale, governed solely by bubble size and independent of membrane or external media. For millimetric bubbles this yields milliseconds, matching $\tau_c$ in Fig.2d. Evacuation times for air and H$_2$ plotted against $R$ collapse on the predicted $R^{3/2}$ scaling (Fig.3c).

\textbf{Mode III.} In Mode II, evacuation enlarges the radius of curvature, lowering Laplace pressure, while the line force increases with decreasing contact angle. The stage ends when the bubble reaches the receding air contact angle, triggering rapid depinning and motion over a few hundred microns at constant angle, followed by re-pinning after ~10 ms. Successive stick–slip motions then complete the collapse within $10–100$ ms (Fig. S8). For high-permeability membranes ($\kappa > 100\,\mu$m$^2$), high inertia can produce a central dimple (Fig. S9), bringing the center into contact before depinning and slightly accelerating the dynamics.

\textbf{Discussion.} The interplay of the three time scales in Mode II is captured by an energy balance for a pinned bubble, $\dot{E}_\gamma \sim \dot{E}_\eta +\dot{E}_{\eta_g} + \dot{E}_i$ where $\dot{E}_\gamma\,\sim\,\gamma R U$ is the Laplace pressure work, $\dot{E}_\eta \sim \eta U^2 R$ is the dissipation in liquid (bulk), $\dot{E}_{\eta_g} \sim \eta_g (U/\phi)^2 L (\phi R^2/ d^2)$  is Poiseuille friction in the pores, and $\dot{E}_i\,\sim\,\rho U^3 R^2$ is bubble inertia. With $U\,\sim\,R/\tau$, the equation for evacuation time reads

\begin{eqnarray}
\tau^2 -\frac{\eta}{\sqrt{\rho \gamma R}} \Big(1+\frac{RL}{d^2 \Phi}\frac{\eta_g}{\eta} \Big) \sqrt{\frac{\rho R^3}{\gamma}}\tau -\frac{\rho R^3}{\gamma} \sim 0\,.
\end{eqnarray}

Nondimensionalizing with the inertio-capillary time, $\sqrt{\rho R^3/\gamma}$, gives $We + Oh(1+1/(Da^*\eta^*)) \sqrt{We} -1 \sim 0$, with $We= \rho RU^2/ \gamma$ is Weber number, $Oh=  \eta/\sqrt{\rho \gamma R}$ Ohnesorge number, $\eta^* =\eta/\eta_g$ the viscosity ratio and $Da^*=d^2 \Phi/RL$ the modified Darcy number ($d^2 \Phi/R^2$ is the classical Darcy number $Da$, corrected here by the ratio of length scales). The coefficient of $\sqrt{We}$, called $\psi$ hereon, captures the interplay of inertial, liquid viscous, and gas viscous dissipation. For small $\psi$ the inertial Rayleigh limit is recovered ($We \sim 1$); for large $\psi$ the viscous limit is obtained ($We \sim {Da^*}^2 {\eta^*}^2/Oh^2$ (Darcy time). Thus, the general solution of Eq. 5 recovers all the three asymptotic limits described earlier.

\begin{figure*}
\includegraphics[width =  0.78\textwidth]{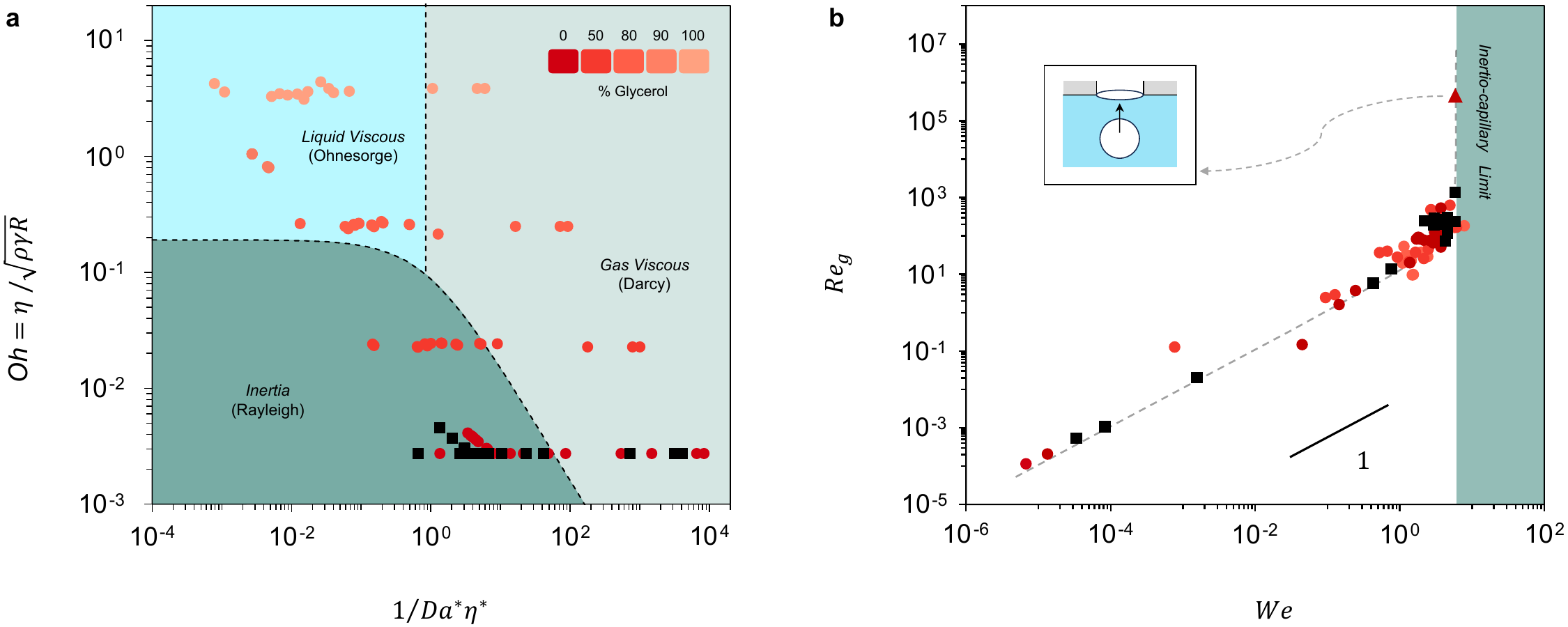}
\caption{\label{fig:wide}  \textbf{Phase Map.}\,(a) Phase map of bubble evacuation showing Rayleigh regime (inertia), Ohnesorge regime (liquid viscosity) and Darcy regime (gas viscosity). The dashed black line separating inertial and viscous regimes follows $Da^* \eta^* = 6.3 Oh/(1-5.3 \cdot Oh)$,  while the line separating liquid and gas viscous regimes corresponds to  $Da^* \eta^* > 0.5$ .  \textbf{b.} The Reynolds number of the gas $Re_g$ when plotted against Weber number, $We$ collapses the entire dataset (excluding glycerol). The dashed line corresponds to $ Re_g = 65\, We/(6-We)$. The green-shaded region indicates the forbidden regime beyond the inertio-capillary limit. The triangular marker on the asymptote corresponds to the free interface limit ($\Phi = 1$).}
\end{figure*}

The separation of the three solutions can be mapped in the  $Oh-{Da^*}^{-1}  {\eta^*}^{-1}$ phase space (Fig 4a). Pre-factors obtained from independent measurements in the Rayleigh (inertial), Ohnesorge (visco-capillary), and Darcy (gas viscous) limits allow quantitative mapping of regime boundaries: the inertial bound follows $Da^* \eta^* > 6.3 Oh/(1-5.3 \cdot Oh)$ (black) and the liquid–gas viscous transition follows $Da^* \eta^* > 0.5$  (red). 

\textit{The Rayleigh-Darcy Regime:} Introducing Reynolds number of gas escaping through pores, $Re_g = \rho (U/\phi) d/\eta_g$, Eq. 4 can be recast for the Rayleigh–Darcy regime, where liquid viscosity is negligible, as
\begin{eqnarray}
Re_g \sim We/(1-We),
\end{eqnarray}

revealing a finite-$We$ singularity. This expresses a fundamental property of curvature-driven debubbling: no matter how quickly the gas evacuates, the inertio-capillary limit, $(\rho R^3/\gamma)^{1/2}$, cannot be surpassed.  

We further note that a nearly identical law (with opposite sign) also emerges in a phase-inverted system of drop impact and ejection from a single-pore sieve \cite{lorenceau2003drops}. Despite the added complexity of debubbling with multiple pores, dissipation pathways, evacuation modes, and pinning, the appearance of the same singularity underscores the universality of the limit set by inertio-capillarity: for drops, this occurs at $We = 3.6$, whereas for bubbles, we find  $We = 6$. 

Remarkably, this minimum-time limit can be reached at a permeability well below that of a free interface $\Phi \approx 1$, revealing a regime where the membrane-laden interface behaves effectively as a 'rigidified' air-liquid interface, as seen in Fig 2d. Matching the Darcy (gas viscous) and Rayleigh (inertial) solutions gives the crossover permeability as $\kappa_i \sim \eta_g L\sqrt{R}/\sqrt{\rho \gamma}  \sim 314\,\mu$m$^2$, attainable with $d = 50 \mu$m, $\Phi \approx 0.13$.

\textit{The Darcy-Ohnesorge Regime:}   For $\eta > \sqrt{\rho \gamma R} \sim 100 $mPa$\cdot$s, inertia is negligible. The Darcy to visco-capillary crossover occurs at $\kappa_\eta \sim \eta_g RL/\eta \sim 10\,\mu$m$^2$, as seen in the glycerol data of Fig. 2d. In this viscous limit, Mode I dewetting slows to the visco-capillary speed $\gamma / \eta$, yet $\tau_I$ remains below the Mode II evacuation time (Fig. S7), so the dynamics is dominated by Mode II.

Finally, the critical permeabilities $\kappa_i$ and $\kappa_\eta$ in conjunction with the phase map provide a framework for appropriate debubbler design that would expedite bubble evacuations, as illustrated in Fig 1a.

%Yet, Fig 1a shows bubbles at the free interface lingering for 1-100s. This raises the natural question: \textit{how does an aerophilic high-permeability membrane differ interfacially from a bare liquid-air interface?}

%The answer to this lies in rigidity of the solid membrane that suppresses the bounces of the bubble, whereas a bare liquid surface can deform (and even stretch) allowing for a significant cushioning of the impact of the approaching bubble, as seen in the timelapse images of Fig.\,5a, b. This sustains the water film separating the bubble from the external media, as well as the bubble itself, for a much longer time – on the order of seconds or more – a thousand times longer than the case of a membrane! However, once ruptured, both a bare interface and a high-permeability membrane show evacuations within $\tau_c= 4.7 \pm 0.6 $ms.  Conversely, a low-permeability membrane can suppress bouncing, but the total evacuation time may extend to several seconds or longer, thereby negating the membrane's effectiveness (Fig. 5c).

\textit{Conclusion:} Aerophilic debubbling, spanning scales from populations to bubbles to pores, unfolds through a multi-scale dynamics in which evacuation is dictated by a pinned contact line mode. Across six decades in Weber number, four in Ohnesorge number, and seven in Darcy number, this mode remains the slowest and rate-limiting step. Its timescale reflects the interplay of three fundamental scales: the Darcy time $\eta_g LR^2/(d^2 \Phi \gamma)$ of gas evacuation, the Ohnesorge time $\eta R / \gamma $ of moving the outer viscous media and a Rayleigh time of $(\rho R^3/\gamma)^{1/2}$ for bubble collapse. The Rayleigh time sets the inertio-capillary limit for the fastest evacuation, conditions uniquely realized by a highly permeable aerophilic membrane that behaves as a 'rigidified' air-liquid interface. Our results reveal a broader design space for interfacial materials and architectures that couple inertia, viscosity, and permeability, offering opportunities to discover new regimes of curvature-driven dynamics across physical, chemical, and biological systems. This motivates further exploration of how interfacial properties are influenced by pore geometry, tortuosity, collective effects, surfactants, stability, and regenerative plastron designs capable of sustaining bubble removal indefinitely.

\textbf{Acknowledgements:} We thank Jack Lake for his help in the fabrication of membranes, and acknowledge the funding of MIT Lincoln Laboratory for this work.

\bibliography{apssamp}% Produces the bibliography via BibTeX.

\end{document}